\documentclass{article}
\usepackage[utf8]{inputenc}

\usepackage{amsmath,amssymb}
\usepackage{amsfonts}
\usepackage{mathrsfs}
\usepackage{graphicx}
\usepackage{authblk}
\usepackage{cite}

\newcommand{\mean}[1]{\left\langle #1\right\rangle}

\newcommand{\braket}[2]{\left\langle\left. #1\right| #2\right\rangle}

\newcommand{\up}[2]{#1^{(#2)}}

\newcommand{\cost}[0]{\mathcal{C}}
\newcommand{\Tr}[1]{\text{Tr}\left(#1\right)}

\begin{document}

\title{Information Transfer as a Framework for Optimized Phase Imaging}

\author[1]{Stewart Koppell}
\author[1]{Mark Kasevich}

\affil[1]{{\small Physics Department, Stanford University, 382 Via Pueblo Mall, Stanford, CA 94305, USA}}

\maketitle

\begin{abstract}
In order to efficiently image a non-absorbing sample (a phase object), dedicated phase contrast optics are required. Typically, these optics are designed with the assumption that the sample is weakly scattering, implying a linear relation between a sample's phase and its transmission function. In the strongly scattering, non-linear case, the standard optics are ineffective and the transfer functions used to characterize them are uninformative. We use the Fisher Information (FI) to assess the efficiency of various phase imaging schemes and to calculate an Information Transfer Function (ITF). We show that a generalized version of Zernike phase contrast is efficient given sufficient foreknowledge of the sample. We show that with no foreknowledge, a random sensing measurement yields a significant fraction of the available information. Finally, we introduce a generalized approach to common path interferometry which can be optimized to prioritize sensitivity to particular sample features. Each of these measurements can be performed using Fourier lenses and phase masks. 
\end{abstract}

\section{Introduction}

In a phase contrast microscope, transparent objects are imaged using optics which convert phase variations to amplitude variations. This modality is important for visible light (notably, biological samples), x-rays\cite{Fitzgerald2000,Bonse1965,Wen2013}, and electrons\cite{Reimer2008}. While some phase contrast is intrinsic in systems with a limited numerical aperture (NA) \cite{Hawkes1994} and more can be generated by adding defocus\cite{Johnson1973}, much of the information about the sample phase shift can only be accessed with dedicated optics. Zernike developed the first method for optically-generated phase contrast using a phase-shifting filter in the backfocal plane of the objective lens (or some conjugate plane) \cite{Zernike1942,Zernike1942_2}. Zernike phase contrast (ZPC) is particularly effective for imaging weak phase objects (WPOs), which have transmission functions close to unity. A probe passed through a WPO will retain a strong undiffracted component which can be used as an interferometric reference. Many phase contrast applications involve phase objects for which the WPO approximation (WPOA) is dubious\cite{Vulovic2014,Misell1976}. For these applications ZPC is only partially effective. When the undiffracted component of the beam is entirely depleted, for example due to a strongly scattering sample matrix, then ZPC produces no contrast at all.

Some common phase contrast methods are compatible with strongly scattering samples, for example the class of schemes sensitive to phase gradients which include Differential Interference Contrast \cite{Allen1969}, Hoffmann Modulation Contrast \cite{Hoffman1977}, and Spiral Phase Contrast \cite{Furhapter05}. However these techniques are insensitive to low spatial frequency features, making them sub-optimal for some measurements. When it is possible to establish a reference channel which circumvents the sample then Quantitative Phase Contrast\cite{Hu2019,Park2018} and various versions of holography\cite{Gabor1949,mollenstedt1956,cowley1992} are possible. These generate some contrast for phase objects of any strength, and their limitations are not as obvious. 

Comparing the effectiveness of these methods is especially difficult outside of the WPOA. In the strong scattering regime the imaging process remains linear with respect to the sample transmission function but becomes non-linear with respect to the sample phase. As a result of this non-linearity, the performance of the imaging system will depend on the joint properties of the optics and the particular sample. One example of an attempt to move beyond the WPOA is Generalized Phase Contrast (GPC)\cite{Gluckstad2009}. GPC, like ZPC, uses the undiffracted probe component as a reference wave. The relative phase and extinction applied to the reference wave can be optimized based on foreknowledge about the sample to maximize the visibility (contrast) or peak irradiance or to establish an unambiguous phase-to-intensity mapping. While GPC avoids invoking the WPOA, it still relies on a strong undiffracted component in the exit wavefunction. To form an even more general theory of phase imaging we must consider a wider class of measurements.

To this end, we recast the imaging process as a many-parameter estimation problem and employ the Fisher Information (FI) to optimize it. While the FI is a prominent tool for experimental design, especially in the field of optics, it is usually applied to optimize measurements of one or a few image parameters. To apply it in a more general imaging scenario with $n$ unknown parameters we must compute the $n^2$ elements of the FI matrix (FIM) describing all of the parameters and their correlations. For even modestly sized images ($n\gtrsim100$) the FIM is expensive to calculate, let alone optimize over all possible measurements. As the optimum may depend on the particular sample, it will be critical to develop efficient heuristics rather than to attempt an explicit optimization for each measurement. Implementing the measurements will require programmable optics. Such technology is available for optical microscopy (i.e. spatial light modulators) and is newly emerging for electron microscopy\cite{Verbeeck2017}.

In the field of quantum metrology, the FI maximized over all measurements permitted by quantum mechanics is called the Quantum Fisher Information (QFI, \cite{Holevo1982,Helstrom1967,Braunstein1994}). The measurement which achieves this maximum generally depends on the values of the parameters being measured. This is no obstacle to many QFI applications, where the goal is to make increasingly precise measurements of an already well-characterized parameter. In the limit of high measurement resources, which we will call the asymptotic regime, we can efficiently measure an unknown parameter by allocating a negligible fraction of the resources for pre-estimation. In situations where measurement resources are limited, which we will call the Bayesian regime, the optimal measurement may depend strongly on the foreknowledge of the parameters \cite{DemkowiczDobrzaski2011}. A measurement sequence in the Bayesian regime will ideally be adaptive so that each measurement is refined using information gathered from previous measurements. Rather than considering the properties of such a measurement sequence, we will focus on optimizing an individual measurement. 

In order to constrain the scope of this project, we make several simplifying assumptions. We assume the sample is a pure phase object with a negligible depth of field, that the measurement is performed with a deterministic source of unentangled scalar particles (i.e. polarization/spin degrees of freedom are not considered), and that the dominant source of noise is projection noise.

In the next section we motivate the transition from contrast to information and introduce the the relevant FI formalism. In section 3 we apply this formalism in the asymptotic regime to the idealized scenario of multi-phase phase estimation (MPE) where arbitrary, lossless transformations can be applied to the exit wavefunction. This perspective helps to clarify the value of a reference channel when projection noise alone limits the measurement efficiency. In section 4 we explore MPE in the Bayesian regime, where foreknowledge of the sample becomes a second limitation. We will develop a suite of methods for phase measurements of strongly scattering samples which are effective for various optimization priorities and levels of foreknowledge. Finally, in section 5 we restrict the optimization to a set of measurements which can be implemented with only a few optical components and take into account the limited NA of the objective lens.

\section{From Contrast to Information}
The properties of a linear optical system can be described by the point spread function or by its Fourier transform, the optical transfer function.  The complex modulus of the optical transfer function is called the modulation transfer function or the Contrast Transfer Function (CTF, especially in electron microscopy\cite{Reimer2008}). The CTF characterizes the frequency-dependent efficiency with which the system transports information from the sample to the detector. For absorbing imaging targets, spatial resolution (limited by lens aberrations and the NA) is often the primary concern. For phase-shifting imaging targets, the CTF expresses another important limitation: how efficiently the optics convert phase variations to intensity variations. 

The CTF is insufficient for describing the properties of the transfer optics when there is not a one-to-one correspondence between spatial frequencies in the sample phase and spatial frequencies in the detected intensity. For example, a strong sinusoidal phase grating, unlike an amplitude grating, diffracts to many orders. We might account for this by replacing the conventional CTF with a scattering matrix (or vector-valued function) which, for each spatial frequency $q_k$ in the sample, gives the resulting intensity contrast at each spatial frequency $q_j$ at the detector. However it is not obvious how condense this data into a figure of merit for optimizing the optics. Instead we will optimize with a cost function which can be constrained with the FI. The FI formalism also provides a quantum limit for the minimal cost which can be used as an optimization benchmark.

We will assume the sample transmission function $\Phi=e^{i\phi}$ can be discretized into $n$ regions with unknown phase shifts $\{\phi_{k}\}_{k=0}^{n-1}$. Let $\Theta=[\theta_0, \theta_1, ..., \theta_{n-1}]$ be a vector of linear parameters describing $\phi$ in the orthonormal basis $\{\up{v}{a}\}_{a=0}^{n-1}$. A linear parameterization of $\phi$ is a \textit{nonlinear} parameterization of $\Phi$:
\begin{equation}
	\Phi_k(\Theta)=\exp\left(i\sum_{a=0}^{n-1}\theta_a\up{v}{a}_k\right)
\end{equation}
 We will often use a `phase grating basis' where $\up{v}{a}$ is a phase grating with spatial frequency $\vec{q}_a$. The phase grating basis is described explicitly in the appendix (section \ref{sec:PhaseGratingBasis}). In order to estimate the values of $\Theta$ based on a measurement outcome $j$ (i.e. detection  at pixel $j$), we use an estimating function (estimator) $\bar{\Theta}(j)$. The optimization of the measurement is defined by minimizing the expected cost
\begin{equation}
\cost=\sum_jI_j(\Theta)C(\Theta,\bar{\Theta}_j)
\end{equation}
 where $C(\Theta,\bar{\Theta}_j)$ is the cost function and $I_j(\Theta)$ is the probability of result $j$ (i.e. the intensity at detector pixel $j$). A standard choice is the quadratic form $C(\Theta,\bar{\Theta}_j)=\left(\bar{\Theta}_j-\Theta\right)^TW\left(\bar{\Theta}_j-\Theta\right)$ where $W$ is a positive semi-definite (often diagonal) weighting matrix which defines the relative priority of reducing the variance of each of the parameters. Using this cost function, the expected cost is
 \begin{equation}
 	\cost=\Tr{W\Sigma_{\bar{\Theta}}}\qquad\Sigma_{\bar{\Theta}} = \mean{\left(\bar{\Theta}-\Theta\right)\left(\bar{\Theta}-\Theta\right)^T}
 \end{equation}
where $\Sigma_{\bar{\Theta}}$ is the covariance matrix for $\bar{\Theta}$. The quadratic cost function is a fairly universal choice when the measurement variance is expected to be small. For larger variances, this cost function does not reflect the periodicity of $\Phi(\Theta)$. We could construct a periodic cost function as in \cite{DemkowiczDobrzaski2011}, however the phase shift is connected to some non-periodic physical property of the imaging target (e.g. the integral of the index of refraction along the optical axis), and it is ultimately this underlying property we want to measure.

Minimizing the expected cost involves choosing optimal transfer optics and simultaneously an optimal estimator. However, with the choice of the quadratic cost function, the Cramer-Rao Bound (CRB) provides a simple way to calculate the lowest achievable variance of any (unbiased) estimator \cite{cramer1999}. The CRB will describe the performance of the optics without specifying the optimal estimator:

\begin{equation}
	\Sigma_{\bar{\Theta}}\geq(N\mathcal{I}(\Theta))^{-1}\qquad \cost\geq\Tr{W(N\mathcal{I}(\Theta)^{-1}}
\end{equation}
where the first inequality is the usual comparison between positive semidefinite matrixes, $N$ is the number of independent measurements, and $\mathcal{I}(\Theta)$ is the Fisher Information Matrix (FIM) for $\Theta$
\begin{equation}
		\mathcal{I}_{a,b}(\Theta)=\sum_j\frac{1}{I_j}\left(\partial_aI_j\right)\left(\partial_bI_j\right)
		\label{eq:FIM}
\end{equation}
where $\partial_a$ is the derivative with respect to $\theta_a$. The diagonal elements $\mathcal{I}_{a,a}\equiv\mathcal{I}_a$ bound the variance for each individual parameter $\theta_a$. In the WPOA, using the phase grating basis, it is possible to show that the FIM is diagonal and its elements are the square of the CTF values. The details of this correspondence are discussed in the appendix (section \ref{sec:C2FI}). Unlike the CTF, the FIM is meaningful even outside the WPOA. In that sense the diagonal of the FIM can naively be thought of as a generalization of the CTF. A more informative transfer function, which takes into account the off-diagonal elements of the FIM, is discussed at the end of this section.

The CRB allows us to bypass the consideration of $\bar{\Theta}$ while optimizing the measurement $T$ implemented by the transfer optics. We can make the dependence of FI on the measurement $T$ explicit by writing $\mathcal{I}(\Theta,T)$. Then the QFI is
\begin{equation}
\mathcal{J}_{a}(\Theta)\equiv \max_T\ \mathcal{I}_{a}(\Theta,T)
\end{equation}
The CRB applied to the QFI is called the quantum CRB (QCRB) \cite{Helstrom1967}. Since the QFI is independent of $T$, it can be considered a measure of the `information' about $\theta_a$ available in the exit wavefunction. To be precise, the QFI describes the variance-reducing power of a measurement and has units of $1/\theta_a^2$ rather than entropy (bits), which is generally considered a more elementary measure of information. Nevertheless, QFI is seen as a fundamental quantity in quantum metrology. 

Multi-parameter measurements are limited by a quantum FIM (QFIM) which is larger than the FIM (in the positive semidefinite sense) for any particular measurement. Whereas the quantum information limit is always attainable in the single parameter case, the matrix bound for multiple parameters may not be attainable when the parameters are associated with incompatible observables \cite{Barndorff_2000}. For example, the limited NA of an objective lens restricts the transverse momentum of the exit wavefunction, thereby performing a counterfactual measurement incompatible with spatial phase measurements of the sample. Even when the QFIM is unattainiable, the cost function can be used to identify an optimal measurement. However the optimal measurement will generally depend on the particular value of $\Theta$. Therein lies the paradox described in the introduction: to construct the optimal measurement, one must first know the result of the measurement. In the Bayesian regime, we can only design measurements to maximize the \textit{expected} FI.

We can express our foreknowledge of $\Theta$ using a probability distribution $\lambda(\Theta)$. For example, the WPO condition could be incrementally relaxed by setting $\lambda(\Theta)=\prod_a\mathcal{N}(\theta_a;\sigma^2)$ where each $\theta_a$ is drawn from an independent normal distribution with zero mean and variance $\sigma^2\ll 1$. A plausible source of foreknowledge is a known diffraction pattern. In this case, we can apply the principle of indifference and assume a uniform distribution over all phase objects with the same diffraction pattern. To sample $\lambda$, we can apply the Gerchberg–Saxton algorithm\cite{Gerchberg72} using the condition that the probe intensity must be uniform at the sample. Given $\lambda$, the goal is to minimize the weighted average of the expected variance. The expected measurement cost is 
  \begin{equation}
\mean{\cost}_\lambda=\Tr{W\mean{\Sigma_{\bar{\Theta}}}_\lambda}
\label{eq:Cfull}
 \end{equation}
where $\mean{.}_\lambda$ is the expectation with respect to $\lambda$. This cost is related to the FI by the van Trees bound \cite{vanTrees2004} (also known as the Bayesian CRB \cite{Gill1995}) 
 \begin{equation}
	  \mathcal{V}(\lambda,T)=\mathcal{I}(\lambda)+N\mean{\mathcal{I}(\Theta,T)}_\lambda\qquad\mean{\Sigma_{\bar{\Theta}}}_\lambda\geq\mathcal{V}^{-1}(\lambda,T)
 \end{equation}
where
\begin{equation}
	\mathcal{I}_{a,b}(\lambda)=\int d^n\Theta \frac{1}{\lambda(\Theta)}\partial_a\lambda(\Theta)\partial_b\lambda(\Theta)
\end{equation} 
 To promote this to a quantum bound, we should maximize $\mean{\mathcal{I}(\Theta,T)}_\lambda$ over all possible measurements. To form a tight upper bound, the maximization should be done after taking the expectation value, which produces the Quantum van Trees Information \cite{Martinez-Vargas2017}. In general, this bound can only be computed by finding the specific measurement which achieves it. Instead we will use generalized QFI (GQFI) $\mathcal{Z}$ \cite{Paris2009,Martinez-Vargas2017} which is obtained by simply replacing $\mean{\mathcal{I}(\Theta,T)}_\lambda$ with $\mathcal{J}$: 
\begin{equation}
	 \mathcal{Z}(\lambda)=\mathcal{I}(\lambda)+N\mathcal{J}\qquad\mean{\Sigma_{\bar{\Theta}}}_\lambda\geq\mathcal{V}^{-1}(\lambda,T)\geq\mathcal{Z}^{-1}(\lambda)
\end{equation}
While this bound is typically unattainable it is generally easier to calculate and thus more suitable as an optimization benchmark. Caution is warranted in interpreting these bounds, as it may not be simple or even possible to devise an efficient estimator (one which saturates the bound) with limited information. We may regard the lower bound on $\mean{C}_\lambda$ as the value we would assign a measurement in retrospect, after collecting enough information to accurately estimate $\Theta$. 
 
When using a uniform weighting $W=\mathbb{I}$, the cost is minimized by prioritizing sensitivity to parameters with large prior variance. This is sometimes undesirable. Suppose the sample consists of a WPO $\Phi(\Theta_f)$ embedded in a strongly scattering matrix $\Phi(\Theta_b)$. The combined transmission function is $\Phi(\Theta)=\Phi(\Theta_f)\Phi(\Theta_b)=\Phi(\Theta_f+\Theta_b)$. We will call $\Theta_f$ the foreground and $\Theta_b$ the background and assume the corresponding prior distributions, $\lambda_f(\Theta_f)$ and $\lambda_b(
\Theta_b)$, are independent so $\lambda(\Theta)=\lambda_f(\Theta_f)\lambda_b(\Theta_b)$. Since $\lambda_b$ contains larger variances, a measurement optimized using cost function defined in Eq. \ref{eq:Cfull} will be tailored for measuring the background. We could attempt to find a non-uniform weighting to increase the cost of foreground error, but it's not obvious how to choose the weights. In the appendix (section \ref{sec:foregroundCost}) we derive a van Trees-like bound on the cost function for variance reduction in the foreground: 
\begin{align}
\mathcal{C}_f&\geq\Tr{W
    \frac{1}{\mathcal{I}(\lambda_f)+N(\mean{\mathcal{I}(\Theta,T)}_{\lambda}-\Delta\mathcal{I})}
    }\label{eq:Cfore}\\
    \Delta\mathcal{I}&=N\mean{\mathcal{I}(\Theta,T)}_{\lambda}\left(\mathcal{I}(\lambda_b)+N\mean{\mathcal{I}(\Theta,T)}_{\lambda}\right)^{-1}\mean{\mathcal{I}(\Theta,T)}_{\lambda}\nonumber
\end{align}
Suppose $\mathcal{I}(\lambda_b)=\eta\mathbb{I}$. In the limit where $\eta\rightarrow0$ (complete ignorance of the background), $\Delta\mathcal{I}\rightarrow\mean{\mathcal{I}(\Theta,T)}_{\lambda}$ so the measurement cost is constant (no information can be gained about the foreground). In the limit where $\eta\rightarrow\infty$ (complete knowledge of the background), $\Delta\mathcal{I}\rightarrow0$ and $\lambda\rightarrow\lambda_f$, so the bound on $\mathcal{C}_f$ becomes identical to the standard van Trees bound. This cost function tends to prioritize sensitivity to parameters which have a small prior background variance. A lower bound on this cost function for all possible measurements is obtained by replacing $\mean{I(\Theta,T)}_{\lambda}$ with $\mathcal{J}$. 

While the cost functions are useful for optimization, they provide little insight into the properties of a particular measurement. For this purpose, it will be useful to define an information transfer function (ITF) which describes the information gained about each parameter. The diagonal of the FIM is not sufficient for this purpose, as it does not account for correlations between parameters, both from the prior distribution and the measurement. A full account of the information gain (in bits) is given by the relative entropy (also known as the KL-divergence) between the prior and posterior probability distributions for $\theta_a$ \cite{kullback1951}. There are two problems with using the relative entropy to define the ITF. First, it would require specifying a rule for updating $\lambda$ for each possible measurement result - in other words specifying an estimator $\bar{\Theta}$. Second, there is no clear way to normalize an ITF based on the relative entropy. If instead the ITF is defined in terms of the FI, then the first problem is solved by the van Trees bound and the second is solved by the limit placed on $\mathcal{V}$ by $\mathcal{Z}$. Therefore we define the ITF as the decrease variance achieved for each parameter (determined using the van Trees bound) relative to the decrease in variance allowed by the GQFI: 
\begin{equation}
 \mathcal{H}(a;\lambda,T)=\frac{\sigma_a^2(\lambda)-(\mathcal{V}^{-1})_{a,a}(\lambda,T)}{\sigma_a^2(\lambda)-(\mathcal{Z}^{-1})_{a,a}(\lambda)}
 \end{equation}
where $\sigma_a^2(\lambda)=\Sigma_{a,a}(\lambda)$ and $\Sigma(\lambda)$ is the covariance matrix for $\lambda$. The maximum value of the ITF is 1, and when $T$ is expected to produce no new information about parameter $\theta_a$, then $\mathcal{H}(a;\lambda,T)=0$. When $\mean{\mathcal{I}}_\lambda$, $\mathcal{J}$, and $\mathcal{I}(\lambda)$ are all diagonal, then the ITF is simply the ratio of the FI and the QFI. When $\Theta$ is expressed in the phase grating basis we will write $\mathcal{H}(\vec{q}_a;\lambda,T)$. As shown in the appendix (section \ref{sec:C2FI}), the ITF is equal to the square of the CTF in the WPOA. Note that unlike a standard transfer function, $\mathcal{H}(\vec{q}_a;\lambda,T)$ depends on spatial frequencies in the sample phase $\phi$ rather than spatial frequencies in transmission function $\Phi$. Also, unlike a linear optical transfer function which describes the properties of the optics alone, the ITF depends on the joint properties of the optics (via $T$) and the particular sample (via $\lambda$). 

We will use a similar formulation to evaluate optimization outcomes in terms of the decrease in cost indicated by the van Trees bound relative to the maximum decrease allowed by the GQFI. For the cost function in Eq. \ref{eq:Cfull},
\begin{equation}
 \Delta C(\lambda,T)=\frac{\Tr{W\Sigma(\lambda)}-\Tr{W\mathcal{V}^{-1}(\lambda,T)}}{\Tr{W\Sigma(\lambda)}-\Tr{W\mathcal{Z}^{-1}(\lambda)}}
\end{equation}

\section{Multi-Phase Estimation with Full Foreknowledge}
\label{sec:MPE}

Before applying the above to phase imaging with optics with limited NA, we will consider the more idealized scenario of multiple phase estimation (MPE) to clarify some of the fundamental limitations of phase imaging with various amounts of foreknowledge. MPE is a well-studied problem in the field of quantum metrology \cite{Humphreys2013,Liu_2019}. Instead of free space modes, the probe states of MPE occupy $n$ discrete channels upon which we can apply arbitrary (lossless) transformations. In order to keep close analogy with phase imaging, we will imagine the channels are arranged in a grid so we may parameterize the $\phi$ in terms of its 2D spatial frequency components. Of course, the actual spatial arrangement of channels in MPE is irrelevant. A reference channel with known phase is generally available and the goal is to find the optimal probe state. We will assume the probe is a pure, single particle state with amplitude $\alpha_j$ in channel $j$ and amplitude $\beta$ in the reference channel ($\sum_j\alpha_j^2+\beta^2=1$). The QFIM for a pure state $\psi$ can be written explicitly\cite{Holevo1982}:
\begin{equation}
\mathcal{J}_{a,b}=4\Re\{\braket{\partial_a\psi}{\partial_b\psi}-\braket{\psi}{\partial_a\psi}\braket{\partial_b\psi}{\psi}\}
\label{eq:QFIM}
\end{equation}
For the case where $n=1$, it is simple to verify that $\mathcal{J}=1$ and a measurement which achieves this limit can be performed with a Mach-Zehnder interferometer (MZI) with $\alpha^2=\beta^2=1/2$. A natural guess for an efficient measurement for $n>1$ is to divide the probe evenly among $n$ parallel MZIs (so half of the total probe intensity still passes through the reference arm). The FIM for this measurement is $\mathcal{I}=\mathbb{I}/n$ so the total information is $\Tr{\mathcal{I}}=1$ and the bound on the total variance from the CRB is $\mathcal{C}=\Tr{\Sigma_{\bar{\Theta}}}\geq \Tr{\mathcal{I}^{-1}}=n^2$. However the quantum limit is superior: $\Tr{\mathcal{J}}=4n/(1+\sqrt{n})^2$ and $\mathcal{C}\geq n(1+\sqrt{n})^2/4$. This implies there is some advantage to simultaneous parameter estimation (if we allow $\psi$ to be a multi-particle entangled state, then this relative advantage is even more pronounced\cite{Humphreys2013}).

In order to explain the advantage of simultaneous estimation, it is helpful use a parameterization which diagonalizes $\mathcal{J}$. If we use a uniform probe $\alpha_j=\alpha$, the QFIM has two distinct parameter eigenspaces. One corresponds to the average phase shift $\theta_0=\frac{1}{n}\sum_j\phi_j$ and has eigenvalue $\mathcal{J}_0=4\alpha^2\beta^2\leq1/n$ which is maximized using probe amplitudes $\beta^2=n\alpha^2=1/2$. The other eigenspace has rank $n-1$ and contains information about all parameters independent of $\theta_0$. Its eigenvalue is $\mathcal{J}_{\perp}=4\alpha^2\leq4/n$, which achieves its largest value when $\beta=0$. Thus, all but one of the degrees of freedom can be measured optimally without a reference channel (see \cite{Goldberg2020} for an analysis of quantum multi-phase estimation without a reference channel), and the total variance is minimized by setting $\beta^2\sim0$ for large $n$ (explicitly, $\beta^2=\sqrt{n}/(n+\sqrt{n})\sim1/\sqrt{n}$). Simultaneous estimation schemes have an advantage, then, because they are able to invest more in the $\mathcal{J}_\perp$ eigenspace, where fewer measurement resources are required to achieve the same variance reduction. 

In some microscopy applications the relevant measurement resource is the total dose, $d=n\alpha^2$. In this case, we should maximize $\mathcal{J}/d$. The matrix eigenvalues become $\mathcal{J}_0=4\beta^2/n$ and $\mathcal{J}_{\perp}=4/n$. This consideration does not change the conclusion that the reference channel is not helpful in the $\mathcal{J}_{\perp}$ eigenspace. Indeed, there are often practical advantages in dispensing with the reference channel. In many imaging applications, the value of $\mean{\phi}$ is irrelevant (e.g. the thickness of the sample matrix) and may even be considered a nuisance parameter. For example, when an imaging system has multiple optical axes, their relative phase stability becomes an added engineering challenge\cite{Mir2012}. We will proceed under the assumption that $\mean{\phi}$ is an extraneous parameter and specialize to measurements which are in-line (lacking a reference channel) and therefore only sensitive to the $\mathcal{J}_{\perp}$ eigenspace (we will suppress the $\perp$ subscript in the future). We will also continue to assume that the probe amplitude is uniform across the channels. With these assumptions, $\mathcal{J}=(4/n)\mathbb{I}$. Since $\mathcal{J}$ is a scalar matrix, it is invariant to reparameterization: a measurement which achieves $\mathcal{J}$ is optimal for estimating any individual parameter or set of parameters. 

\section{Multi-Phase Estimation with Limited Foreknowledge}

We will now discuss several types of in-line measurements which are useful with various levels of foreknowledge. We will assume the measurements are projective so that $I_j=|T_{j,k}\Phi_k|^2$ for some unitary matrix $T$ (this precludes measurement schemes which use multiple detectors to make non-commuting measurements). We can factorize $T$ as
\begin{equation}
    T=U^*MU
\end{equation}
where $U$ (a unitary matrix) is the measurement eigenbasis and $M$ is a diagonal matrix of unit-norm eigenvalues $M_{q,q}=e^{i\mu_q}$. A sufficient condition for $T$ to achieve the QFIM limit is if $U$ concentrates all of the intensity in the exit wavefunction into a single eigenvector. This is possible in the asymptotic regime (where $\Phi(\Theta)$ is known) by setting $U=\mathcal{F}\Phi^{-1}$ where $\mathcal{F}$ is the Fourier transform matrix and $\Phi^{-1}$ is the inverse of the sample transmission function. Then its simple to show $\mathcal{I}(\Theta,T)=\mathcal{J}$ if $\mu_{q=0}=\pi/2$ and $\mu_{q>0}=0$. For WPOs ($\Phi^{-1}\sim 1$), this measurement is equivalent to ZPC. The reparameterization-invariance of $\mathcal{J}$ implies that ZPC performs optimally for measuring any feature of a WPO. It is interesting to note that while a general projective measurement of a state with $n$ degrees of freedom is described by a unitary transform with $n^2$ real parameters, an optimal measurement can be performed using only using ZPC optics (with no degrees of freedom) and a phase mask with $n$ degrees of freedom. This makes it practical to design efficient phase imaging optics for any sample using relatively few optical elements.

 \begin{figure}[htbp]
\begin{center}
\includegraphics[width=\linewidth]{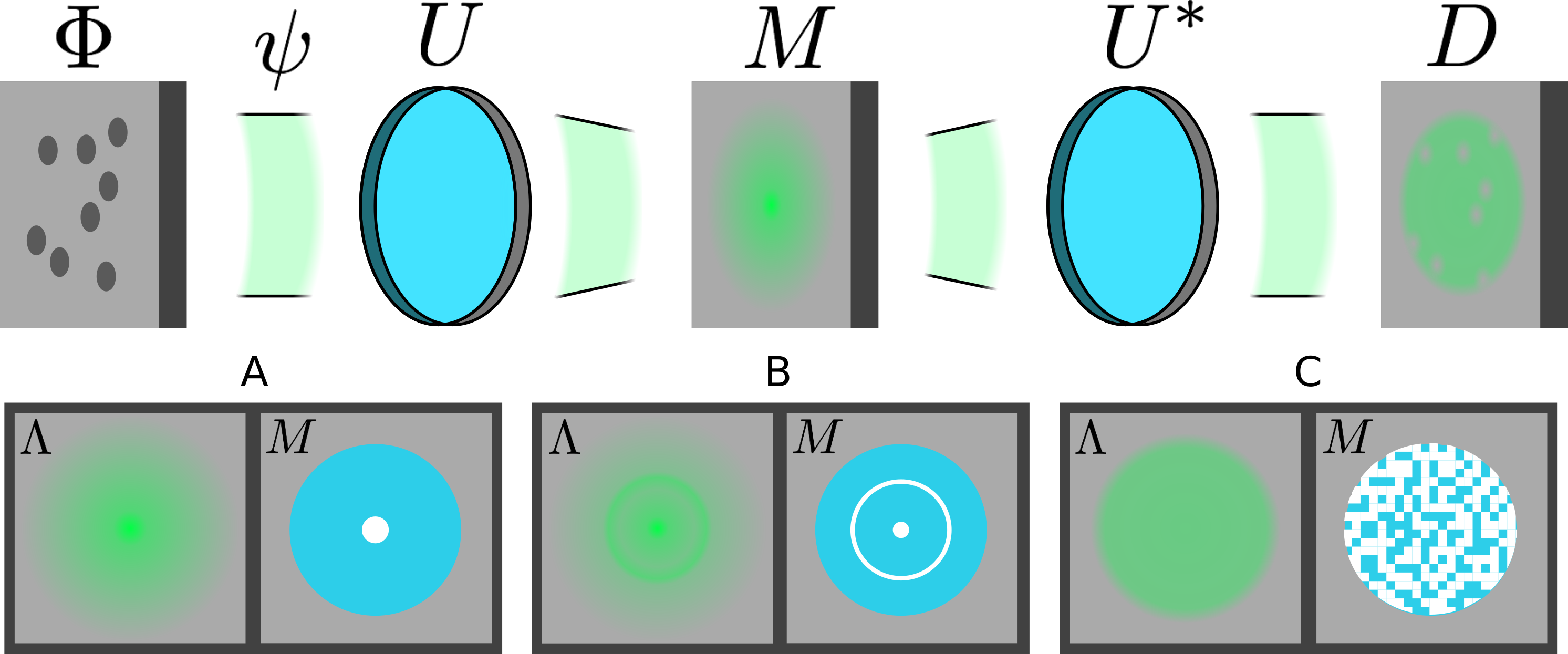}
\caption{Exit wavefunction $\psi$ is formed by passing a uniform, single particle probe through an unknown phase object $\Phi$. A unitary operator $T=U^*MU$ is applied to $\psi$ before measurement at detector $D$. If it is possible to choose a measurement eigenbasis $U$ which condenses $\psi$ into a small subspace, then an efficient measurement can be performed using Generalized Common Path Interferometry (GCPI). A set $Q$ of eigenvectors is designated as the in-line reference. The measurement eigenvalues are then set to $M_q=\exp(i\mu)$ for $q\in Q$ and $M_q=1$ otherwise. The membership of $Q$ and the value of $\mu$ are optimized based on the expected intensity $\Lambda_q$ carried by eigenvector $q$. As a rule of thumb, $Q$ includes the eigenvectors carrying the highest intensities. When $\Lambda_q$ is uniform, $Q$ includes half of the eigenvectors at random. A, B, and C are schematics of example intensity patterns $\Lambda_q$ and the corresponding optimal $M$. The white region represents the set $Q$.}
\label{MPEvsPI}
\end{center} 
\end{figure}

In the Bayesian regime we do not have precise knowledge of $\Phi$ and therefore cannot choose a measurement eigenbasis which concentrates $\psi$ into single eigenvector. As shown in the appendix (section \ref{sec:Projective}) this precludes finding a projective measurement which achieves the quantum information limit. To find an efficient measurement we must optimize based on the prior distribution $\lambda$. As a starting point, we can set $U=\mathcal{F}e^{-i\mean{\phi}_\lambda}$, $\mu_{q=0}=\pi/2$ and $\mu_{q=0}=0$. This measurement is effective when $\lambda$ contains mainly translation-specific prior information, or, equivalently, when the covariance matrix for $\lambda$ is nearly diagonal for the parameterization $\theta_a=\delta_{a,k}\phi_k$. An implementation of this measurement is described in \cite{Juffmann2020}. With little or no translation-specific information, the measurement is ineffective. This can be partially ameliorated by optimizing the phase shift $\mu_{q=0}\equiv\mu$, an approach we will call Generalized ZPC. The maximum efficiency of GZPC will depend on how much intensity $\Lambda_q=\mean{\left|(U\psi)_q\right|^2}_\lambda$ is focused into measurement eigenvector $q=0$. In the appendix (section \ref{sec:ZernikeFI}) we calculate $\mean{\mathcal{I}}_\lambda$ for GZPC in the particular case that each of the parameters are independently, normally distributed with variance $\sigma^2$. For $\Lambda_0=e^{-\sigma^2}\gtrsim0.8$, the result  is $\mean{\mathcal{I}}_\lambda\sim\Lambda_0\mathcal{J}$. This linear approximation underestimates $\mean{\mathcal{I}}_\lambda$ in the region $0.8>\Lambda_0>0.5$, where it plateaus to a value of $\mean{\mathcal{I}}_\lambda\sim\frac{3}{4}\mathcal{J}$. For $\Lambda_0<0.5$, $\mean{\mathcal{I}}_\lambda$ drops precipitously. The effectiveness of GZPC can be extended to lower values of $\Lambda_0$ by increasing the value of $\mu$, in which case $\mean{\mathcal{I}}_\lambda\sim\frac{3}{4}\mathcal{J}$ is maintained until $\Lambda_0<\frac{1}{4}$.
 
If no foreknowledge of $\Phi$ is available, we can assign a uniform distribution to each phase $\phi_k$. In this case $\Lambda_0\sim1/n$ and GZPC (for any $\mu$) is uninformative. However if we randomly set each $\mu_{q}$ to either 0 or $\pi$, the expected Fisher Information is $\mean{\mathcal{I}}_\lambda\sim\frac{1}{2}\mathcal{J}$. This measurement, which we will call random sensing, is similar to the technique described by Oe and Namura \cite{Oe18} which uses a diffuser to generate in-line phase contrast. Oe and Namura rely on the WPOA to reconstruct the phase object. For strongly scattering samples, we must resort to a general phase retrieval algorithm such as Gerchberg-Saxton or Fienup \cite{Fienup82}. Despite the factor of 2 discrepancy between the FI for random sensing and the QFI, this measurement out-performs the parallel Mach-Zehnder interferometer scheme (provided that the reconstruction algorithm produces the full variance reduction allowed by the CRB).

In some circumstances, it is possible to find a measurement more efficient than both GZPC and random sensing. For example, if there exists a set $Q$ of measurement eigenvectors with $|Q|\ll n$ such that $R=\sum_{q\in Q}\Lambda_q\sim1$, then it's simple to show that setting $\mu_{q\in Q}=\pi/2$ and $\mu_{q\not\in Q}=0$ defines a measurement which gives $\mean{\mathcal{I}}_\lambda\sim\mathcal{J}$. Such a set exists, for example, if $\Phi$ is a crystal with a known, sharp diffraction pattern (but perhaps unknown translation). In general, if it is possible to choose $U$ which concentrates $\psi$ into a small subspace, then this strategy produces an efficient measurement: if $|Q|\ll n$ and $R\sim1$, $\Tr{\mean{\mathcal{I}}_\lambda}\sim\Tr{\mathcal{J}}R(1-|Q|/n)$. However, rather than trying to explicitly optimize $U$, we will focus on applications where the nature of the foreknowledge about the sample leads to a natural choice of $U$. For example, $U=\mathcal{F}$ is the natural choice when $\lambda$ is induced by an expected diffraction envelope. More generally, we will assume $\Lambda$ is given for a particular $U$ and proceed to optimize $M$.
 
Having chosen the measurement eigenbasis, we can define a family of measurements called Gemeralized Common Path Interferometry (GCPI) which is parameterized by the set $Q$ and the phase shift $\mu$ applied to measurement eigenvectors $q\in Q$. A procedure for optimizing over $\mu$ and $Q$ is described in the appendix (section \ref{sec:Optimization}). The optimization is especially likely to identify a measurement more efficient than GZPC or random sensing when specializing to foreground variance reduction (using the cost function \ref{eq:Cfore}) or to high-spatial frequency measurements. For example, in dose-limited electron microscopy, the high spatial frequency features degrade quickly and the achievable resolution scales with the fourth power of the dose \cite{DeJonge2018}. This motivates a parameter weighting $W_{a,a}=|\vec{q}_a|^4$.

\begin{figure}[htbp]
\begin{center}
\includegraphics[width=\linewidth]{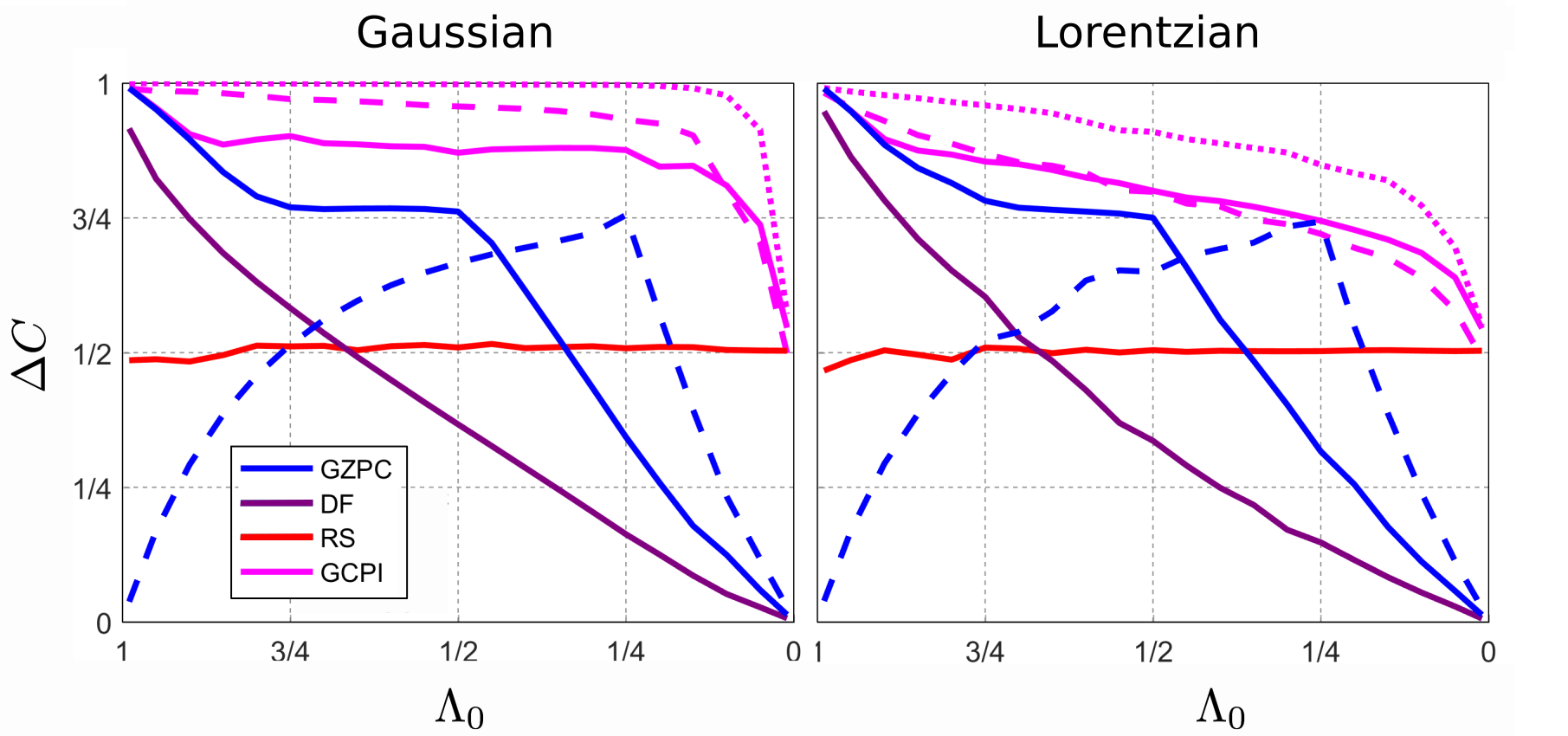}
\caption{In-line multi-phase estimation schemes for $n=16^2$ channels and prior distributions induced by Gaussian (left) and Lorentzian (right) diffraction envelopes with unscattered intensity $\Lambda_0$. The quantity $\Delta C$ is the average reduction in the weighted variance relative to the limit imposed by the GQFI for various phase contrast schemes: generalized Zernike phase contrast (GZPC) with $\mu=\pi/2$ (solid line) and $\mu=\pi$ (dashed line), dark field (DF), random sensing (RS), and generalized common path interferometry (GCPI). GCPI is optimized for the total variance (solid line) or just the foreground variance (dashed and dotted lines) and for a uniform weighting (dashed line) or for $W_{a,a}=|\vec{q}_a|^4$ (solid and dotted lines).}
\label{fig:AVR}
\end{center}
\end{figure}

In Fig. \ref{fig:AVR}, we compare the efficiency of GZPC, random sensing, GCPI, and dark field microscopy. On the left, $\Delta C$ is calculated for samples which are known to have a Gaussian intensity distribution $\Lambda$ in some basis $U$ for various peak intensities $\Lambda_0$. When the weighting on the parameters is uniform and the optimization is done using the full cost (Eq. \ref{eq:Cfull}), GCPI offers no advantage over the best choice among the other methods. However, when the weighting is $W_{a,a}=|\vec{q}_a|^4$, GPCI achieves a lower measurement cost by sacrificing sensitivity at low spatial frequencies in exchange for increased sensitivity at high spatial frequencies. GCPI also exploits this trade-off to out-perform other methods when specializing to foreground variance reduction using the cost function Eq. \ref{eq:Cfore}. On the right, $\Lambda$ is a 2D Lorentz distribution $\Lambda(\vec{q})=\frac{1}{2\pi}\frac{w}{\left(|\vec{q}|^2+w^2\right)^{3/2}}$ with $w=(2\pi\Lambda_0)^{-1/2}$. In accordance with the rule of thumb described above, GCPI is especially effective for the Gaussian prior, where $\Lambda$ is more concentrated. 

\section{Phase Imaging with a Limited Numerical Aperture}

Given some foreknowledge of the diffraction pattern of a sample, the measurements described in the previous sections can be performed using a spatial phase modulator to implement $M$ and two Fourier lenses to implement $U$ and $U^*$. Given some translation-specific foreknowledge, it may also be beneficial to add a second spatial phase modulator to a conjugate-image plane before the first lens to implement $U=\mathcal{F}e^{-i\mean{\phi}_\lambda}$. In this section we will account for loss due to the limited numerical aperture of real lenses. This adds some intrinsic phase contrast which becomes significant for strongly scattering samples, but also reduces the total amount of information which can reach the detector. The achievable efficiency will depend on how well the probe, which is focused in the condenser aperture to provide plane wave illumination, can be refocused in a conjugate plane after passing through the sample.

Let $A(|\vec{q}|)$ be a hard aperture function with $A(|\vec{q}|<q_\text{max})=1$ and $A(|\vec{q}|>q_\text{max})=0$. For a weak phase object (or an amplitude object), $A$ blocks all information about spatial frequencies with a magnitude larger than $q_{\text{max}}$. However the intensity pattern at the detector depends on \textit{all} spatial frequencies present in a strong phase object, regardless of $q_{\text{max}}$. For example, the diffraction pattern of the superposition two phase gratings at spatial frequencies $\vec{q}_a$ and $\vec{q}_b$ contains the beat frequencies $\vec{q}_a\pm\vec{q}_b$. Even if both $|\vec{q}_a|>q_\text{max}$ and $|\vec{q}_b|>q_\text{max}$, it's possible that $|\vec{q}_a-\vec{q}_b|<q_\text{max}$. This principle makes it possible to achieve superresolution using structured illumination\cite{Lukosz1966}. Since the illumination is also limited by the NA, structured illumination can only improve resolution over the standard limit by a factor of two. But with a sufficiently informative prior distribution $\lambda$ providing known structure in the sample itself, diffraction no longer imposes a fundamental resolution limit \cite{Lukosz1966,Guerra1995}. There remains, however, an information limit. 

The measurements which can be applied to exit wavefunction $\psi$ using diffraction-limited optics are non-projective and cannot be described using a unitary transfer function of rank $n$. However we will assume that the measurement applied to the wavefunction exiting the Fourier aperture $\Psi_q=A_q\mathcal{F}(\psi)_q$ is unrestricted. Then the diffraction-limited QFIM, $\tilde{\mathcal{J}}$, can be calculated by applying Eq. \ref{eq:QFIM} to $\Psi$. We will neglect the small amount of additional information available when using a deterministic source (i.e. the FI associated with the total intensity missing at the detector). Unlike the QFIM for MPE, $\tilde{\mathcal{J}}$ depends on $\Theta$ and is not diagonal, making the calculation of the GQFI less trivial: $\tilde{\mathcal{Z}}(\lambda)=\mathcal{I}(\lambda)+\mean{\tilde{J}(\Theta)}_\lambda$. The off-diagonal elements are generally small and a good approximation is
\begin{equation}
    \tilde{\mathcal{J}}_{a,b}(\Theta)\approx4\delta_{a,b}\braket{\partial_a\Psi}{\partial_a\Psi}=4\delta_{a,b}\sum_qA_q\left|\mathcal{F}(\up{v}{a}\cdot\psi)_q\right|^2
\end{equation}
While more strongly scattering samples send a larger portion of the probe intensity outside the NA, they are also more sensitive to spatial frequencies higher than $q_{\text{max}}$ through the beating effect described above. These effects act in equal measure and the fractional QFI lost to the aperture $\Tr{\mathcal{J}-\tilde{\mathcal{J}}}/\Tr{\mathcal{J}}\sim\sum_qA_q/n$ is roughly constant regardless of $\lambda$.

\begin{figure}[htbp]
\begin{center}
\includegraphics[width=\linewidth]{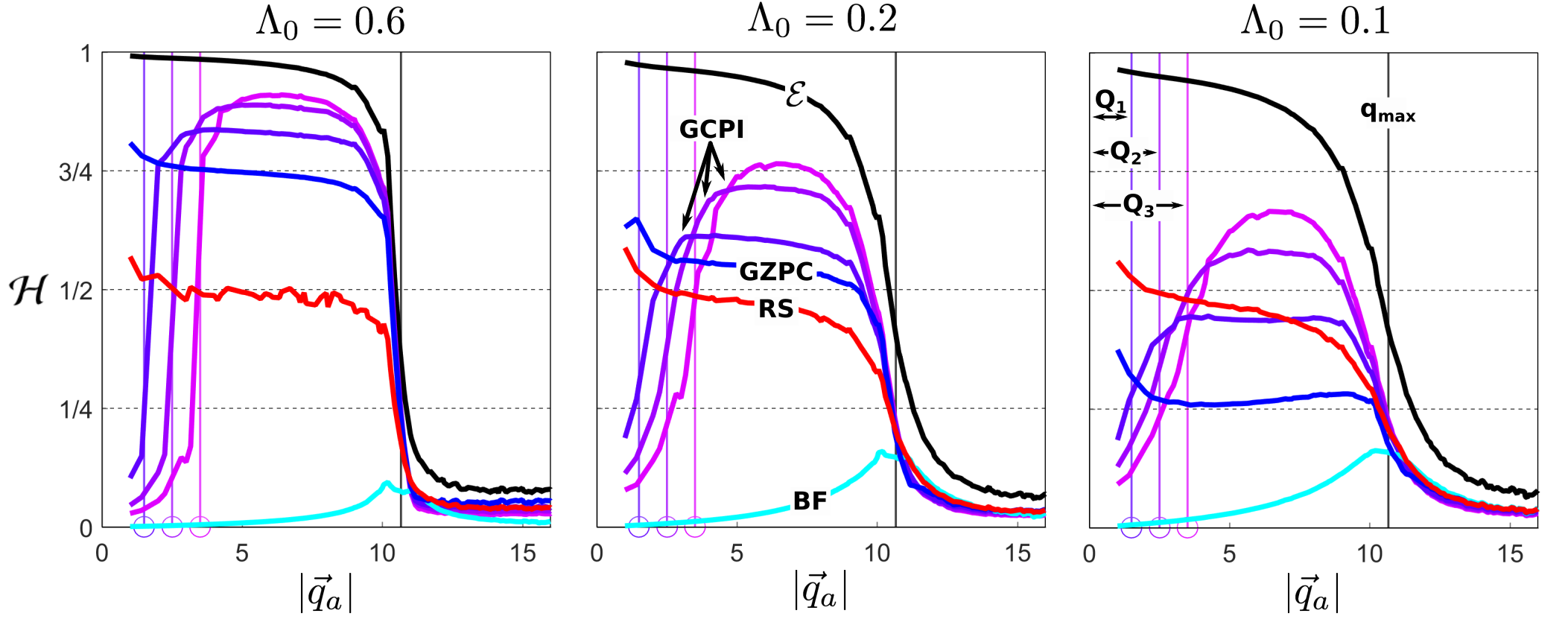}
\caption{Information Transfer Function (ITF) for various phase imaging schemes for a prior distribution $\lambda$ induced by a Lorentzian diffraction envelope with unscattered intensity $\Lambda_0=0.6$ (left), $\Lambda_0=0.2$ (middle), and $\Lambda_0=0.1$ (right). The horizontal axis is the magnitude of the spatial frequency in the sample phase. The black vertical line at $|\vec{q}_a|=q_{\text{max}}$ marks the largest spatial frequency in the exit wavefunction allowed through the Fourier plane aperture. The black envelope labeled $\mathcal{E}$ is the information limit set by the aperture. The other curves are the ITF for Generalized Zernike Phase Contrast (GZPC, using $\mu=\pi/2$ for $\Lambda_0=0.6$ and $\mu=\pi$ for $\Lambda_0=0.2$, 0.1), random sensing (RS), bright field (BF), and Generalized Common Path Interferometry (GCPI) with phase shift $\mu=\pi/2$ applied a successively larger sets of Fourier coordinates $Q_1$, $Q_2$, and $Q_3$.}
\label{fig:ITF_PI}
\end{center}
\end{figure}

Using $\tilde{\mathcal{Z}}(\lambda)$, we can write an envelope function for the ITF which represents the maximum diffraction-limited variance reduction
\begin{equation}
    \max_{T}\ \mathcal{H}(|\vec{q}_a|;\lambda,T)\leq\mathcal{E}(|\vec{q}_a|,\lambda)=\frac{\sigma_a^2(\lambda)-\tilde{\mathcal{Z}}_{a,a}^{-1}(\lambda)}{\sigma_a^2(\lambda)-\mathcal{Z}^{-1}_{a,a}(\lambda)}
    \label{eq:ITFenvelope}
\end{equation}
Fig. \ref{fig:ITF_PI} shows the ITF for various phase contrast schemes. Since natural images often have spectra with power $\sim q^{2}$ \cite{Burton1987,Field1987,tolhurst1992}, we assume the diffraction pattern has a 2D Lorentz distribution with unscattered intensity $\Lambda_0=0.6$ (left), $\Lambda_0=0.2$ (middle), and $\Lambda_0=0.1$ (right). The black curve is the envelope function defined in Eq. \ref{eq:ITFenvelope}. Comparing the three plots, we see that as more intensity scatters outside the NA, the decrease in $\mathcal{E}$ below $q_{\text{max}}$ is accompanied by an approximately equal increase above $q_{\text{max}}$. The cyan curve is the ITF for the intrinsic (bright field) contrast due to scattering outside the NA. The blue curve is the ITF for GZPC using the optimal phase ($\mu=\pi/2$ for $\Lambda_0=0.6$, $\mu=\pi$ for the other two). The red curve is the ITF for random sensing. The remaining curves are ITFs for GCPI using $\mu=\pi/2$ with varying $|Q|$. As $|Q|$ increases, information about high spatial frequency parameters is gained at the cost of information about low spatial frequency parameters. While the decrease in low spatial frequency information is strictly a disadvantage from the perspective of any (positively weighted) cost function, it may be a positive feature in some circumstances. For example, filtering out low spatial frequencies may simplify data interpretation (finding an efficient estimator).

\begin{figure}[htbp]
\begin{center}
\includegraphics[width=\linewidth]{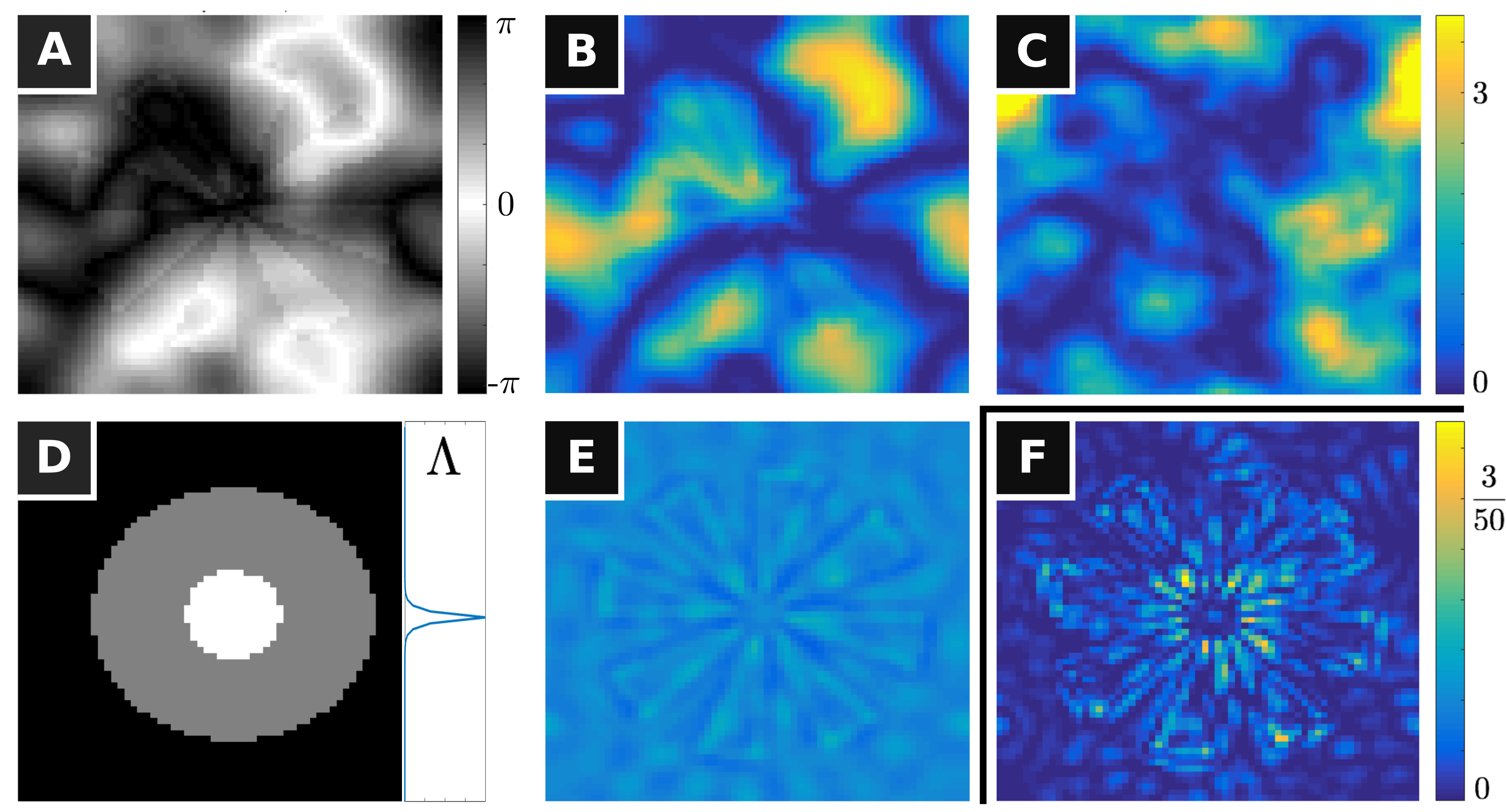}

\caption{Phase contrast imaging simulation of a weak phase object (a 20 $\mu$m diameter pinwheel with phase thickness  $\pi/10$) embedded in a strong phase background with a Lorentzian power spectrum. The background scatters 80\% of the $\lambda=500$nm plane-wave illumination. A: The sample phase shift in radians. B: The detected intensity pattern using ZPC with a $\pi$ phase shift. C: The detected intensity pattern using random sensing. D: The Fourier-plane filter for GCPI  optimized for foreground detection. The black region lies outside the NA. The grey and white regions are completely transmissive and have a relative phase shift of 0.52$\pi$ rad. A cross section of the expected diffraction intensity $\Lambda$ is shown with the same vertical scale. E: The detected intensity pattern using the GCPI filter shown in D. F: The detected intensity pattern using the filter shown in D but with the transmissivity central region set to 0.}
\label{fig:WPO}
\end{center}
\end{figure}

In Fig. \ref{fig:WPO} we optimize a GCPI filter for measuring a WPO in a strongly scattering background using the cost function in Eq. \ref{eq:Cfore}. We again assume a Lorentzian diffraction pattern and set $\Lambda(\vec{q}=0)=0.2$. The foreground WPO is a 20 $\mu$m diameter pinwheel. The phase of the combined foreground and background is shown in (A). The detected intensity distributions using ZPC with $\mu=\pi$ (B), random sensing (C), and GCPI (E) are shown with identical color scales. The optimized Fourier filter for GCPI is shown in (D). A phase shift of $\sim0.52\pi$ is applied in the central (white) region relative to the outer (grey) region. The black region is absorptive and establishes a NA of 0.8 using 500nm light. Besides providing good contrast for high spatial frequency features, foreground-optimized GCPI filters out the much of the background. A similar filtering affect can be achieved simply by blocking the prominent spatial frequencies in the background. In (F) the GCPI filter is modified so that the central (white) region is completely absorbing. This high-pass filter produces significantly less contrast: the color scale in (F) is emphasized by a factor of 50 compared to the color scale in (E).

\section{Conclusion}
Outside of the WPOA, each spatial frequency in the sample phase affects many spatial frequencies in the intensity at the detector. This non-linearity makes it difficult to design efficient phase imaging transfer optics. We have approached this problem using FI as a rigorous optimization framework and developed an information transfer function to study the properties of various measurements of pure phase objects. As a rule of thumb, the amount of information that can be extracted from a single measurement depends on how well the exit wavefunction can be concentrated into a small subspace using foreknowledge of the sample. GZPC is a family of measurements described by a single parameter which can be optimized for efficient phase imaging if at least 20\% of the probe intensity can be refocused. When GZPC is ineffective, a random sensing measurement can be employed without any optimization at the cost of complicating the measurement interpretation. A third option, GCPI, performs at least as well as GZPC and random sensing and is especially effective when specialized to measuring high spatial frequencies or imaging WPOs in a strongly scattering background. It would be straightforward to extend these methods and the ITF to phase objects with finite depth of field and finite absorption, and also to include lens aberrations and limited coherence. The ITF could also be used to characterize an aggregate measurement including multiple modalities (e.g. phase contrast and fluorescence) by summing their individual FIMs.

\section{Acknowledgements}
This work was supported by the Gordon and Betty Moore Foundation and the Department of Energy grant DE-SC0019174-00

\appendix
\section{The Phase Grating Basis}
\label{sec:PhaseGratingBasis}
We want a spatial frequency parameterization $\phi$. Since $\phi$ is not imaginary, we should not use a discrete Fourier basis. Instead, we could a discrete cosine or sine transform. But these contain `half frequency' elements. It will be more convenient to parameterize $\phi$ in a basis where every element has a simple Fourier representation. Inspired by the real discrete Fourier transform\cite{Ersoy1985}, we choose

\begin{equation}
	\up{v}{a}(\vec{r})=
	\begin{cases}
	\sqrt{\frac{2}{n}}\cos(2\pi \vec{q}_a\cdot\vec{r}) & a\in A\\
	\sqrt{\frac{2}{n}}\sin(2\pi \vec{q}_a\cdot\vec{r}) & a\not\in A
	\end{cases}
\end{equation}

which has Fourier components

\begin{equation}
	\up{\tilde{v}}{a}(\vec{r})=
	\begin{cases}
	\frac{1}{\sqrt{2}}(\delta_{\vec{q}_a}+\delta_{-\vec{q}_a}) & a\in A\\
	\frac{1}{\sqrt{2}}(\delta_{\vec{q}_a}-\delta_{-\vec{q}_a}) & a\not\in A
	\end{cases}
\end{equation}
where $A$ contains $a$ if $\vec{q}_a=-\vec{q}_{a}$ and otherwise contains one of either $a$ or $b$ such that $\vec{q}_b=-\vec{q}_{a}$.

\section{The Information Transfer Function}
\label{sec:C2FI}
The definition of contrast most often used to calculate the CTF is the Michelson contrast
\begin{equation}
C^M=\frac{I_{\max}-I_{\min}}{I_{\max}+I_{\min}}
\end{equation}
Suppose the sample is a phase grating with spatial frequency $q$ and amplitude $\delta\theta_q\ll1$. The CTF can be written
\begin{equation}
	\mathcal{C}(q)=C^M/\delta\theta_q
\end{equation}
As an example, suppose we use a plane wave probe and the transfer function $T$ applies a phase shift $\mu(q)$ to each Fourier component $q$. Then $I=|T(\psi)|^2=|\mathcal{F}\{\mathcal{F}\{\psi\}e^{i\mu}\}|^2$ where $\mathcal{F}$ represents the action of a Fourier lens. Using the WPOA, we find the CTF is
\begin{equation}
	\mathcal{C}(q)=2|\sin(\mu(q)-\mu_0)|
\end{equation}
In some formulations, the CTF is a signed quantity with negative contrast indicating dark fringes. The CTF is also often normalized so that $|\mathcal{C}(q)|\leq1$.

Consider a phase object built from a superposition of phase gratings with amplitudes $\Theta=[\theta_{1}, \theta_{2},...,\theta_{n}]$. If we perturb one of these amplitudes by  a small amount $\delta\theta$, how much contrast will that perturbation generate? The answer using the Michelson contrast depends on the intensities measured by only two detector pixels (at the locations of $I_{\text{max}}$ and $I_{\text{min}}$). Clearly this summary statistic is too coarse-grained to capture the full effect of the perturbation. As an alternative, we can define the CTF using the root-mean square Weber contrast
\begin{equation}
	\mathcal{C}(q)=\sqrt{\sum_j\left(\frac{1}{\theta_q}C^W_j(q)\right)^2}
\end{equation}
where $C^W$ is the Weber contrast
\begin{equation}
	C^W_j(q)=\frac{I-I_b}{I_b}
\end{equation}
where $I_b=I\big |_{\delta\theta_q=0}$. These definitions of the CTF are entirely equivalent in the WPOA. Now consider the  square of the  Weber CTF for small perturbations $\delta\theta_q\rightarrow 0$:
\begin{align}
 \lim_{\delta\theta_q\rightarrow0}\mathcal{C}^2(\theta_q)&=\sum_j\left(\frac{1}{\theta_q}\frac{I-I_b}{I_b}\right)^2\\
 &=\sum_j\left(\frac{1}{I_b}\frac{\partial}{\partial\theta_q}I_b\right)^2\\
 &=\mathbb{E}\left[\left(\frac{\partial}{\partial\theta_q}\log(I_b)\right)^2\right]=\mathcal{I}_q
\end{align}
where $\mathbb{E}$ is the expectation value. The final expression is the definition of the FI for parameter $\theta_q$. We can also show this equivalence by calculating the diagonal elements of the FIM for a WPO using the phase grating basis and unitary transfer function $T$:

\begin{align}
	\mathcal{I}_{a,a}(\Theta=0,T)&=\sum_j\frac{4}{|T(\psi)|^2_j}\Re\left\{\overline{T(\psi)}_j\partial_aT(\psi)_j\right\}^2\\
	&=4/n\sin^2(\mu(\vec{q}_a)-\mu(\vec{q}_0))
\end{align}
which (apart from the normalization factor $1/n$) is the square of the CTF. However, we cannot interpret the diagonal of the FIM as a transfer function outside of the WPOA, as the off-diagomal elements may be important. Nevertheless, it will be useful to formulate a transfer function based on the FIM to help visualize the properties of a particular measurement. 
We define the information transfer function (ITF) for a measurement $T$ as the ratio of the maximum variance reduction for parameter $\theta_a$ achievable by $T$ (as determined by the van Trees bound) to the maximum variance reduction for parameter $\theta_a$ allowed for any measurement (as determined by the GQFI). In general, we write the function as $\text{ITF}(|\vec{q}_a|;\Theta,T)$ assuming $\Theta$ is expressed in the phase grating basis and that $T$ and $\lambda$ respect radial symmetry around $\vec{q}=0$. Explicitly, the ITF is

\begin{equation}
    \text{ITF}(|\vec{q}_a|;\Theta,T)=\frac{\sigma_{a}^2(\lambda)-(\mathcal{V}^{-1})_{a,a}(\lambda,T)}{\sigma_{a}^2(\lambda)-(\mathcal{Z}^{-1})_{a,a}(\lambda)}
\end{equation}

For large $n$, $\mean{I_a(\Theta,T)}_\lambda\leq\mathcal{J}_a=4/n\ll1$ and if $\sigma_a^2(\lambda)\ll n/4$, we can expand $\mathcal{V}^{-1}$ and $\mathcal{Z}^{-1}$ in powers of $\sigma_a^2(\lambda)\mean{I}_\lambda$  and $\sigma_a^2(\lambda)J$, respectively:

\begin{align}
    (\mathcal{V}^{-1})_{a,a}(\lambda,T)/\sigma_a^2(\lambda)&=1-\sigma_a^2(\lambda)\mean{\mathcal{I}_a(\Theta,T)}_\lambda+\mathcal{O}\left(\sigma_a^4(\lambda)\mean{\mathcal{I}_a(\Theta,T)}^2_\lambda\right)\\
    (\mathcal{Z}^{-1})_{a,a}(\lambda)/\sigma_a^2(\lambda)&=1-\sigma_a^2(\lambda)\mathcal{J}_a+\mathcal{O}\left(\sigma_a^4(\lambda)\mathcal{J}_a^2\right)
\end{align}
and the ITF  becomes
\begin{equation}
    \text{ITF}(|\vec{q}_a|;\Theta,T)\sim\mean{\mathcal{I}_a(\Theta,T)}/\mathcal{J}_a
\end{equation}
This approximation is accurate, for example, in the WPOA, in which case the ITF is the square of the CTF as shown above.

\section{Fisher Information for WPO in a Strong Background}
\label{sec:foregroundCost}
The standard formulation of the cost function measures the expected average variance. An optimized measurement will prioritize sensitivity to the parameters with the largest prior variances. Suppose the sample consists of a WPO (the foreground) embedded in a strongly scattering, unknown background. Let parameter vector $\Theta_f=[\theta_{f;0},\theta_{f;1},...]$ with prior distribution $\lambda_f(\Theta_f)$ describe the foreground and $\Theta_b=[\theta_{b;0},\theta_{b;1},...]$ with prior distribution $\lambda_b(\Theta_b)$ describe the background, so the total transmission function is
\begin{equation}
    \Phi(\Theta_f)_k\Phi(\Theta_b)_k=\exp\left(i\sum_{a=0}^{n-1}\left(\theta_{f;a}+\theta_{b;a}\right)\up{v}{a}_k\right)
\end{equation}
We cannot separately measure $\theta_{f;a}$ and $\theta_{b;a}$, but we can adjust the cost function to specifically reward reduction of the foreground variance.

Let $\lambda_{\text{tot}}$ be the prior distribution for $\Theta_{\text{tot}}=[\Theta_f,\Theta_b]$. The covariance matrix for the estimator of the combined parameter vector is constrained by the van Trees bound
\begin{equation}
    \mean{\Sigma_{\hat{\Theta}_{\text{tot}}}}_{\lambda_{\text{tot}}}\geq\frac{1}{\mathcal{I}(\lambda_{\text{tot}})+N\mean{\mathcal{I}(\Theta_{\text{tot}})}_{\lambda_{\text{tot}}}}
    \label{eq:vTBtot}
\end{equation}
The cost function 
\begin{equation}
    \mean{C}_{\lambda_{\text{tot}}}=\Tr{W_{\text{tot}} \mean{\Sigma_{\hat{\Theta}_{\text{tot}}}}_{\lambda_{\text{tot}}}}
\end{equation}
is equivalent to the standard cost function when
\begin{equation}
    W_{\text{tot}}=\frac{1}{2}
    \begin{pmatrix}
W & 0 \\
0 & W
\end{pmatrix}
\label{eq:EvenWeighting}
\end{equation}
but can be specialized to foreground variance reduction using 
\begin{equation}
    W_{\text{tot}}=
    \begin{pmatrix}
W & 0 \\
0 & 0
\end{pmatrix}
\end{equation}
We can write $\mean{\mathcal{I}(\Theta_{\text{tot}})}_{\lambda_{\text{tot}}}$ as a $2\times2$ block diagonal matrix, where each block is $\mean{\mathcal{I}(\Theta)}_{\lambda_{\text{tot}}}$. We will also write $\mathcal{I}(\lambda_{\text{tot}})$ in block form 
\begin{equation}
    \mathcal{I}(\lambda_{\text{tot}})=
    \begin{pmatrix}
\mathcal{I}(\lambda_{\text{tot}})_{11} & \mathcal{I}(\lambda_{\text{tot}})_{12} \\
\mathcal{I}(\lambda_{\text{tot}})_{12}& \mathcal{I}(\lambda_{\text{tot}})_{22}
\end{pmatrix}
\end{equation}
so that the right hand size in Eq. \ref{eq:vTBtot} is
\begin{equation}
        \begin{pmatrix}
\mathcal{I}(\lambda_{\text{tot}})_{11}+N\mean{\mathcal{I}(\Theta)}_{\lambda_{\text{tot}}} & \mathcal{I}(\lambda_{\text{tot}})_{12}+N\mean{\mathcal{I}(\Theta)}_{\lambda_{\text{tot}}} \\
\mathcal{I}(\lambda_{\text{tot}})_{12}+N\mean{\mathcal{I}(\Theta)}_{\lambda_{\text{tot}}} & \mathcal{I}(\lambda_{\text{tot}})_{22}+N\mean{\mathcal{I}(\Theta)}_{\lambda_{\text{tot}}}
\end{pmatrix}^{-1}
\end{equation}
since the weight matrix has three zero quadrants, we need only calculate the upper left quadrant of this matrix inverse to find
\begin{align}
    \mean{C'}_{\lambda_\text{tot}}\geq&\text{Tr}\bigg(W
    \Big(\mathcal{I}(\lambda_{\text{tot}})_{11}+N\mean{\mathcal{I}(\Theta)}_{\lambda_{\text{tot}}}\\
    &-\left(\mathcal{I}(\lambda_{\text{tot}})_{12}+N\mean{\mathcal{I}(\Theta)}_{\lambda_{\text{tot}}}\right)\left(\mathcal{I}(\lambda_{\text{tot}})_{22}+N\mean{\mathcal{I}(\Theta)}_{\lambda_{\text{tot}}}\right)^{-1}\left(\mathcal{I}(\lambda_{\text{tot}})_{12}+N\mean{I(\Theta)}_{\lambda_{\text{tot}}}\right)
    \Big)^{-1}\bigg)
\end{align}
If we assume the prior distributions for $\Theta_f$ and $\Theta_b$ are  independent so $\lambda_{\text{tot}}=\lambda_f(\Theta_f)\lambda_b(\Theta_b)$, then $\mathcal{I}(\lambda_{\text{tot}})_{11}=\mathcal{I}(\lambda_f)$, $\mathcal{I}(\lambda_{\text{tot}})_{12}=0$, $\mathcal{I}(\lambda_{\text{tot}})_{22}=\mathcal{I}(\lambda_b)$, and
\begin{equation}
    \mean{C_f}_{\lambda_\text{tot}}\geq\Tr{W
    \frac{1}{\mathcal{I}(\lambda_f)+N\mean{I(\Theta)}_{\lambda_{\text{tot}}}-N^2\mean{\mathcal{I}(\Theta)}_{\lambda_{\text{tot}}}\left(\mathcal{I}(\lambda_b)+N\mean{\mathcal{I}(\Theta)}_{\lambda_{\text{tot}}}\right)^{-1}\mean{\mathcal{I}(\Theta)}_{\lambda_{\text{tot}}}}
    }
\end{equation}
As $\mathcal{I}(\lambda_b)\rightarrow\infty$ (meaning the background is known and $\lambda_{\text{tot}}\rightarrow\lambda_f$), this cost function approaches the standard van Trees bound for $\lambda_f$. When $\mathcal{I}(\lambda_b)$ is small (i.e. $\mathcal{I}(\lambda_b)\ll N\mean{\mathcal{I}(\Theta)}_{\lambda_{\text{tot}}}$)  then
\begin{equation}
        \mean{C_f}_{\lambda_\text{tot}}\geq\Tr{W
    \frac{1}{\mathcal{I}(\lambda_f)+\mathcal{I}(\lambda_b)\left(1-\frac{1}{N}\mean{I(\Theta)}^{-1}_{\lambda_{\text{tot}}}\mathcal{I}(\lambda_b)\right)}
    }
\end{equation}
As an example, suppose $\lambda_f$ is independently and identically distributed for each of the parameters so that the prior information matrix is $\mathcal{I}(\lambda_f)=\frac{1}{\sigma^2}\mathbb{I}$. Also suppose $W=\mathbb{I}$ and $\lambda_b(\Theta_b)=\prod_{a=0}^{n-1}\lambda_a(\theta_a)$ where each $\lambda_a$ is normal with zero mean and variance $\sigma_a^2\gg N\sigma^2$.
\begin{align}
    \mean{C'}_{\lambda_\text{tot}}&\gtrsim (n-1)\sigma^2+\frac{\sigma^4}{N}\sum_{a=0}^{n-1}\left(\sigma_a^{-4}\mean{\mathcal{I}_a(\Theta)}_{\lambda_{\text{tot}}}-\sigma^{-2}_a\right)
\end{align}
This cost function has a strong preference for measuring parameters with small $\sigma_a$, where the foreground is more `visible' despite the background. For comparison, if we use the weighting in Eq. \ref{eq:EvenWeighting} we get the standard cost
\begin{align}
    \mean{C}_{\lambda_{\text{tot}}}\gtrsim\sum_{a=1}^{n-1}\frac{1}{\sigma^2+\sigma_a^{-2}+N\mean{\mathcal{I}_a(\Theta)}_{\lambda_{\text{tot}}}}
\end{align}
which gives priority to increasing $\mean{\mathcal{I}_a(\Theta)}_{\lambda_{\text{tot}}}$ for parameters with large $\sigma_a$.

\section{Projective Measurements in the Bayesian Regime}
\label{sec:Projective}
In multi-phase estimation with $n$ = 2 phases, the phase difference $\phi_2-\phi_1$ can be optimally measured without a reference channel or any prior knowledge of the phases using a 50-50 beam splitter. For $n>2$ pixels, the phase differences between neighboring channels can be measured using a series of beam splitters and 2$n-1$ detectors. This measurement is impractical for phase imaging, where $n$ is large and space is limited. Instead, we will consider only projective measurements which can be represented by a unitary matrix $T$ of rank $n$. Here we give an informal argument that projective measurements generally cannot achieve the QFIM in the Bayesian regime.

Using Eq. 5 we can write the FI for parameter $\theta_a$, 
\begin{equation}
	\mathcal{I}_a(\Theta,T)=4\sum_j|T(\partial_a\psi)_j|^2\sin^2(\up{\gamma}{a}_j-\gamma_j)
\end{equation} 
where $\gamma_j=\text{arg}\{T(\psi)_j\}$ and $\up{\gamma}{a}_j=\text{arg}\{T(\up{v}{a}\psi)_j\}$. We can easily see that $\mathcal{I}_a(\Theta,T)=\mathcal{J}_a=4/n$ if and only if $\sin^2(\up{\gamma}{a}_j-\gamma_j)=1$ for all $j$ where $|T(\partial_a\psi)_j|^2>0$. This is possible only if $\Lambda=|U\psi|^2$ has no overlap with $\up{\Lambda}{a}=|U\partial_a\psi|^2$, and the QFIM can only be achieved if this condition is met for all values of $a$. $\Lambda$ can be thought of as the reference component and $\up{\Lambda}{a}$ as a signal component. Achieving the QFIM requires the reference to be completely isolated from the signal. It takes $n-1$ channels to carry information about $n-1$ independent parameters. This limits $\Lambda$ to a single channel. In the Bayesian regime, we will not have sufficient prior information to find a measurement basis where $\Gamma$ occupies a single channel.

\section{Generalized ZPC in the Bayesian Regime}
\label{sec:ZernikeFI}
\begin{figure}[htbp]
\begin{center}
\includegraphics[width=\textwidth]{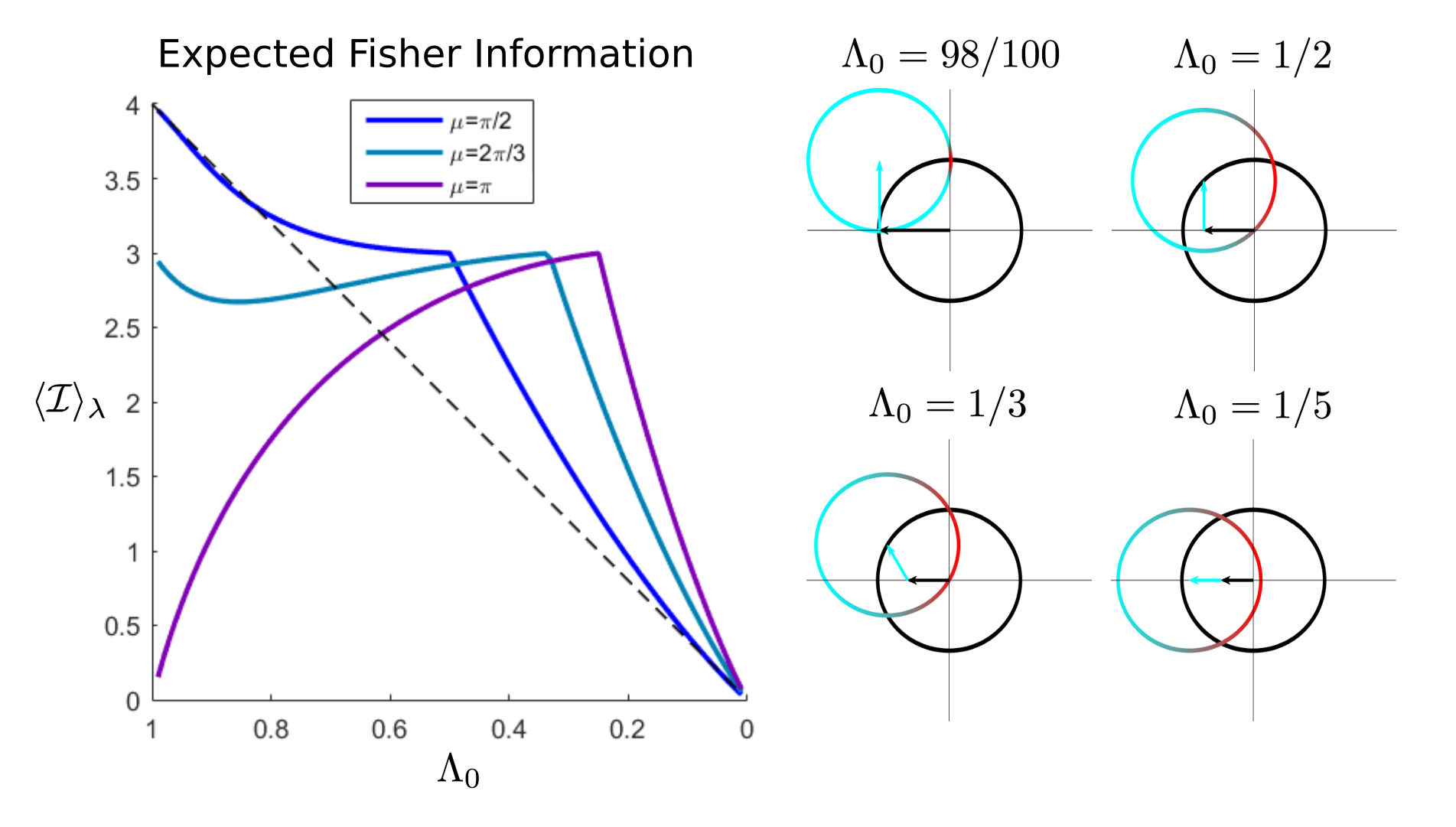}

\caption{Left: Expected Fisher Information $\mean{\mathcal{I}}_\lambda$ for Zernike phase contrast with a prior $\lambda$ which is an independent Gaussian distribution with variance $\sigma^2$ for each phase. The maximum FI and the ideal Zernike phase shift, $\mu$, depend on the unscattered intensity $\Lambda_0=e^{-\sigma^2}$. The dashed black line is 4$\Lambda_0$, which  is a good approximation for $\mean{\mathcal{I}}_\lambda$ when $\Lambda_0>0.8$. Right: Phasor diagrams which show the the action of the transfer optics on the exit wavefunction (represented by the black unit circle). The cyan circle represents the possible values of the wavefunction at the detector, and the red portion represents the probability distribution of the wavefunction. The black and cyan vectors have length $\Lambda_0$ and relative angle $\mu$. For $.25<\Lambda<.5$, the optimal $\mu$ causes the cyan circle to pass through the the origin. For $\Lambda_0>.5$ and $\Lambda_0<0.25$, the optimal values for $\mu$ are $\pi/2$ and $\pi$, respectively.}
\label{fig:phasors}
\end{center}
\end{figure}

Here we calculate the FI for GZPC for a particular, simple prior distribution $\lambda$. The effect of the ZPC optics is to add a phase shift $\mu$  to the zero-frequency component (mean) of the wavefunction exiting the sample. The transfer function can be written
\begin{equation}
    T(\psi_j)=\psi_j+(e^{i\mu}-1)\mean{\psi}
\end{equation}
We will assume the probe amplitude is uniform and the sample is a pure phase object, so $|\psi_j|^2=1/n$. The sample phase is described by n-1 parameters in the vector $\Theta$ which weigh phase grating basis elements $\up{v}{a}$ (we exclude $\theta_0$, which determines the average phase thickness). If the prior distribution on $\Theta$ is $\lambda(\Theta)$, then the expected FI is
\begin{align}
    \mean{\mathcal{I}_{a,b}(\Theta)}_\lambda&=\int d^{n-1}\lambda(\Theta)\sum_{j=1}^n\frac{1}{I_j}\partial_aI_j\partial_bI_j\\
    &=\frac{16}{n}\int d^{n-1}\lambda(\Theta)\sum_{j=1}^n\frac{\sin^2(\mu/2)\Lambda_0|\up{v}{a}_j||\up{v}{b}_j|\cos^2(\phi_j-\phi_0-\mu/2)}{1+4\sin^2(\mu/2)\Lambda_0+4\sin(\mu/2)\sqrt{\Lambda_0}\sin(\phi_j-\phi_0-\mu/2)}
\end{align}
where $\Lambda_0=|\mean{\psi}|^2$ and $\phi_0=\arg{\mean{\psi}}$. Note $\phi_j$ and $\phi_0$ depend on $\Theta$. If we assume that $\lambda$ is independently and identically distributed for each parameter, then we can also write identical and independent distributions $\lambda(\phi_j)=\lambda(\phi)$ for each $\phi_j$. For large $n$, the distribution for $\phi_0$ is narrow (with variance $\sim\frac{1}{n}$) around a mean which we will assume, without loss of generality, is zero. Then
\begin{align}
    \mean{\mathcal{I}_{a,b}(\Theta)}_\lambda&=\delta_{a,b}\frac{16}{n}\sin^2(\mu/2)\Lambda_0\int d\phi\lambda(\phi)\frac{\cos^2(\phi-\mu/2)}{1+4\sin^2(\mu/2)\Lambda_0+4\sin(\mu/2)\sqrt{\Lambda_0}\sin(\phi-\mu/2)}
\end{align}
We can now optimize the Zernike phase, $\mu$, for a particular distribution $\lambda(\phi)$. Suppose $\lambda(\phi)=\frac{1}{\sigma\sqrt{2\pi}}e^{-\phi^2/2\sigma^2}$. The ideal choice of $\mu$ depends on $\Lambda_0=e^{-\sigma^2}$:
 \begin{equation}
	\mu=
	\begin{cases}
	\pm\pi/2 &\Lambda_0\geq\frac{1}{2}\\
	\pm2\arcsin\left(\frac{1}{2\sqrt{\Lambda_0}}\right) & \frac{1}{2}>\Lambda_0>\frac{1}{4}
	\end{cases}
\end{equation}
The expected FI for $\mu=\pi/2$, $2\pi/3$, and $\pi$ are shown in Fig. \ref{fig:phasors}. The figure also shows phasor diagrams which may provide some intuition for the optimal values of $\mu$. 

\section{Details of Numerical Calculations}
\label{sec:Optimization}.

In order to optimize GCPI, the value of $\mu$ and the membership of $Q$ must be jointly optimized based on the prior distribution $\lambda$. When $\lambda$ is induced by an expected intensity pattern $\Lambda$ we sample from $\lambda$ using the Girchberg-Saxton algorithm with a uniform random initial phase distribution. In order to determine $\mathcal{I}(\lambda)$, we estimate the covariance matrix for $\lambda$, $\Sigma_\lambda$, then set $\mathcal{I}(\lambda)=\Sigma_\lambda^{-1}$.

The number of possible sets of $Q$ is combiniatorially large. In order to reduce the complexity of optimizing $Q$, we estimate the value $V_q$ of including $q\in Q$, then set $Q=\{q|V_q\geq  V_*\}$ and optimize the threshold value $V_*$. The estimated value will depend on the cost function. To minimize the weighted average of the expected variance using cost function from Eq. 7, we use 
\begin{equation}
V_q=\frac{\Lambda_q}{\sum_aW_{a,a}\up{\Delta}{a}_{\text{max}}\mean{\up{\Lambda}{a}_q}_\lambda}
\label{eq:value}
\end{equation}
where $\up{\Delta}{a}_{\text{max}}$ is the maximum variance reduction allowed by the GQFI for $\theta_a$. The denominator is the weighted average of the signal components expected to be carried by eigenvector $q$, and estimates the opportunity cost of losing sensitivity to $q$. When optimizing for foreground variance reduction using the cost function from Eq. 11, we use
\begin{equation}
V_q=\frac{\Lambda_q}{\sum_aW_{a,a}\left(\up{\Delta}{a}_{\text{max}}\right)^{-1}\mean{\up{\Lambda}{a}_q}_\lambda}
\label{eq:foregroundValue}
\end{equation}
which has higher value when $\up{\Delta}{a}_{\text{max}}$, the potential reduction in the background variance, is smaller. In many cases, especially when $\Lambda_q$ decreases monotonically with $q$, the same value ranking is obtained simply using $V_q=\Lambda_q$. The optimization proceeds by alternating between minimizing the cost with respect to $\mu$ use Matlab's fminbnd (with $\pi/2<\mu<\pi$), and then minimizing with respect to $V_*$. 

\bibliography{references}
\bibliographystyle{unsrt}

\end{document}